\documentclass[epj-spec]{svjour}
\usepackage{graphics}

\usepackage{amsmath,amssymb}

\usepackage[cmtip,arrow]{xy}
\usepackage{pb-diagram,pb-xy}
\usepackage{pst-plot, pst-coil, pstricks}
\begin{document}
\title{Medium dependence of multiplicity distributions in MLLA}
\author{N\'estor Armesto
  \and Carlos Pajares\and Paloma Quiroga-Arias}
\institute{Departamento de F\'{\i}sica de Part\'{\i}culas and IGFAE,
Universidade de Santiago de Compostela\\15706 Santiago de Compostela, Spain}

\abstract{
We study the modification of the multiplicity distributions in MLLA due to the presence of a QCD medium. The medium is introduced though a multiplicative constant
($f_{med}$) in the soft infrared parts of the
kernels of 
QCD evolution equations. Using the asymptotic ansatz for quark and gluons mean multiplicities
$\langle n_G\rangle=e^{\gamma y}$ and $\langle n_Q\rangle=r^{-1}e^{\gamma y}$ respectively, we study two cases: fixed $\gamma$ as previously considered in the literature, and 
fixed $\alpha_s$.
We
find opposite behaviors of the dispersion of the multiplicity 
distributions with increasing $f_{med}$ in both cases. For fixed $\gamma$ the dispersion decreases, while for fixed $\alpha_s$ it increases.
}

\maketitle

\section{Introduction}\label{intro}

The standard explanation for the phenomenon of jet quenching - the suppression of particles with large transverse momentum produced in nucleus-nucleus collisions compared with the expectations from an incoherent superposition of nucleon-nucleon ones - observed at the Relativistic Heavy Ion Collider (RHIC) at Brookhaven National Laboratory \cite{rhic}, is radiative energy loss (see the reviews \cite{rel}). This explanation implies a modification of the standard QCD radiation pattern (see e.g. \cite{Armesto:2007dt} for a recent proposal).

\begin{figure}
\begin{center}
\psset{unit=0.7cm}
\begin{pspicture}(-2,-1)(8,3)
\psline[linestyle=dashed,linecolor=gray](-2,0)(0,0)
\put(-1.8,0.5){$Q$}
\put(-1.8,-0.5){$E_{jet}$}
\psline[linestyle=dashed](0,0)(2.5,1.9)
\psline[linestyle=dashed](0,0)(2.5,-1.9)
\put(0,0){\line(1,0){7}}
\put(1.5,0){\vector(1,0){0.8}}
\pscoil[coilheight=1.2,coilwidth=0.2,coilarm=0](3,0)(6,1.2)
\psarc(3,0){1.5}{0}{22}
\put(5,0.2){$\theta$}
\put(6.2,0.9){$E_g$}
\end{pspicture}
\end{center}
\vskip 0.5cm
\caption{Kinematical variables in the process of gluon radiation from a parent parton.}
\label{fig0}
\end{figure}
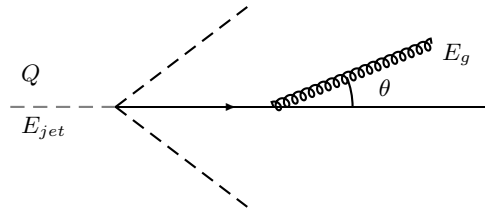

In standard  perturbative QCD, the process of gluon radiation suffers from several types of singularities at each order in $\alpha_s$, see Fig. \ref{fig0}, leading to single and double logarithms, which have to be resummed. 
More specifically, there can be two different singularities when a gluon is emitted: collinear singularities when the gluon emission angle is very small ($\theta\to 0$) leading to $\ln \theta$; and infrared divergences when the emitted gluon takes a very small fraction $x$ of the energy of the parent, leading to $\ln(1/x)$.
These singularities give problems of convergence, thus the need of resummations. There are different resummation schemes, namely: 
\begin{itemize}
\item Leading Logarithmic Approximation (LLA): it resums the single logs $\left[\alpha_s\ln\left(k_T^2/\mu^2\right)\right]^n$, with $k_T$ the transverse momentum of the emitted gluon with respect to the parent parton.

\item Double Logarithmic Approximation (DLA): resummation of collinear and infrared singularities $[\alpha_s\ln(1/x)\ln\theta]^n\sim {\cal O}(1)$.

\item Single Logarithms (SL): takes into account the emission of hard collinear gluons ($\theta\to 0$), $[\alpha_s\ln\theta]^n\sim{\cal O}(\sqrt{\alpha_s})$.

\item Modified Leading Logarithmic Approximation (MLLA):  SL correction to DLA,
\\ $\left[\alpha_s\ln(1/x)\ln\theta+\alpha_s\ln\theta\right]^n\sim [{\cal O}(1)+({\cal O}(\sqrt{\alpha_s})].$
\end{itemize}

The standard radiation pattern in QCD \cite{basics} has been extensively tested by measurements in high energy $e^+ e^-$ and $pp (p\bar{p})$  collisions. While the DLA resummation scheme is known to overestimate the cascading process, as it ignores the energy-momentum conservation since the radiating particle does not loose any energy after soft gluon radiation,  it has been found a very good agreement between the data from OPAL and TASSO and the  MLLA approximation, while TeVatron data require further refinements \cite{Arleo:2007wn}.

When a medium is created in a heavy ion collision, the radiating particle will travel through it. A distortion in the radiation pattern will appear. In this work, we focus on the multiplicity distributions within a jet. We will use a simple medium implementation within MLLA as proposed in \cite{BorghiniWiedemann}, to describe the QCD radiation pattern in medium. We go beyond previous studies \cite{dremin2} by considering different possibilities to fix the parameters in the calculation, which produces qualitivavely different results.

\section{Calculation and results}\label{sec:1}

The computation of multiplicity distributions in QCD has been discussed many times, see e.g. \cite{basics,Dokshitzer:1993dc,dremin}. Here we simply indicate the main formulae:
The multiplicity distributions at a scale (rapidity) $y=\ln (p\theta/Q_0)=\ln (2Q/Q_0)$, with $Q_0$ an
infrared cut-off, is the probability of finding $n$ objects in the final state radiated within an angle $\theta$ from a parton with momentum $p$: 
\begin{equation}\label{eq.1}
P_n(y)=\frac{\sigma_n}{\sum_{n=0}^{\infty}\sigma_n}=\frac{\sigma_n}{\sigma_{inel}}\ .
\end{equation}
%
%
%
%
%
The generating functional is defined in the following way:
\begin{equation}\label{eq.4}
G(y,z)=\sum_{n=0}^{\infty}P_n(y)(1+z)^n.
\end{equation}
%

The normalized factorial moments of range $q$ are defined as
\begin{equation}\label{eq.5}
M_q=\frac{1}{\langle n \rangle ^q}\left.\frac{d^q G(y,z)}{d z^q}\right|_{z=0}=\frac{\sum_{n}P_n n(n-1)...(n-q+1)}{\left(\sum_n P_n n\right)^q}\ ,
\end{equation}
so the generating functional can be expressed as a function of those moments:
\begin{equation}\label{eq.6}
G(y,z)=\sum_{q=0}^{\infty}\frac{z^q}{q!}\langle n\rangle^qM_q\ .
\end{equation}

The normalized factorial moment of range one is equal to 1.
The  normalized factorial moment of range two,
\begin{equation}
M_2=\frac{\langle n(n-1)\rangle}{\langle n \rangle ^2}\ ,
\end{equation}
is related to the normalized dispersion of the distribution,
\begin{equation}
D^2_{norm}=\frac{\langle n^2\rangle -\langle n\rangle ^2}{\langle n\rangle ^2}\ ,
\end{equation}
in the following way:
\begin{equation}\label{eq.7}
M_2=D_{norm}^2+1-\frac{1}{\langle n \rangle}\ .
\end{equation}

The master evolution equation for  splitting of a parton A into partons B and C, with momentum fractions $x$ and $(1-x)$ respectively, is \cite{basics}
\begin{equation}\label{eq.8}
\frac{d G_A(y,z)}{d y}=\frac{1}{2}\sum_{B,C}\int_0^1 d x \frac{\alpha_s(k_T^2)}{2\pi}\Phi_A^{BC} [G_B(y+\ln x ,z ) G_C(y+\ln(1-x),z)-G_A(y,z)].
\end{equation}

The different possibilities for splitting [their respective probabilities] are:
$g{}\to g{(x)} g{(1-x)}$ [$\Phi_G^{GG}(x)$], $g{}\to q{(x)} \bar{q}{(1-x)}$ [$\Phi_G^{QQ}(x)$],  $q{}\to q{(x)} g{(1-x)}$ [$\Phi_Q^{QG}(x)$] and $q{}\to g{(x)} q{(1-x)}$ [$\Phi_Q^{GQ}(x)=\Phi_Q^{QG}(1-x)$].
The evolution equations for the generating functionals of gluon and quark jets in MLLA read
\begin{eqnarray}\label{eq.9}
G'_G(y,z)&=&\int^{1}_{0}dx K^{G}_{G}(x)\gamma_0^2[G_G(y+\ln x,z)G_G(y+\ln(1-x),z)-G_G(y,z)]\\\nonumber
&+&n_f\int^{1}_{0}dx K^{Q}_{G}(x)\gamma_0^2[G_Q(y+\ln x,z)G_Q(y+\ln(1-x),z)-G_G(y,z)],\\
G'_Q(y,z)&=&\int^{1}_{0}dxK^{G}_{Q}(x)\gamma_0^2[G_G(y+\ln x,z)G_Q(y+\ln(1-x),z)-G_Q(y,z)],\label{eq.9.1}
\end{eqnarray}
where the prime denotes the derivative with respect to $y$, and $\gamma_0^2=2 N_c \alpha_s/\pi$, with $N_c=3$ the number of colors and $n_f(=3)$ the number of active flavors.
The negative terms in both equations take into account the probability of no radiation and yield the corresponding integral finite without the need of any regulation of the splitting kernels.

In the absence of any medium, the splitting kernels $K_i^j$ are the standard DGLAP kernels e.g. used for jets in $e^+e^-$ annihilation.
We introduce the medium in the splitting kernels by multiplying the infrared-divergent parts by a constant $f_{med}$ as done in \cite{BorghiniWiedemann}: 
\begin{eqnarray}\label{eq.11}
K_G^{G}(x)&=&\frac{1+f_{med}}{x}-(1-x)[2-x(1-x)],\\\nonumber
K_G^{Q}(x)&=&\frac{1}{4N_c}\left[x^2+(1-x)^2\right],\\
K_Q^{G}(x)&=&\frac{C_F}{N_c}\left[\frac{1+f_{med}}{x}-1+\frac{x}{2}\right],\nonumber
\end{eqnarray} 
where $C_F=(N_c^2 -1)/(2 N_C)$, and $K_A^B$ represents the splitting of a particle $A$ with momentum $p$ into a particle $B$ with momentum fraction $x p$ and another particle $C$ with momentum fraction $(1-x)p$. For $f_{med}=0$ we obviously recover the standard DGLAP kernels. The constant $f_{med}$ grows with increasing medium length and transport coefficient, and decreasing parent-parton energy, but the quantitative relation has not been established yet.

Now we proceed by substituting \eqref{eq.6} into  \eqref{eq.9} and \eqref{eq.9.1}, and by collecting terms with equal powers of the auxiliary variable $z$ in both sides of the equations. In this way we will obtain equations for the mean multiplicities at order $z$, and for the corresponding normalized factorial moments of range two at order $z^2$.

\subsection{First order equations:  ${\cal O}(z)$}

At lowest order in $z$ we get:
\begin{eqnarray}\label{eq.12,13}
\langle n_G(y)\rangle '&=&\int^1_0
dx\gamma_0^2 K^G_G(x) [\langle n_G(y+\ln x)\rangle+\langle
n_G(y+\ln (1-x))\rangle-\langle n_G(y)\rangle]\\\nonumber
&+&n_f\int^1_0 dx\gamma_0^2 K^Q_G(x)[\langle n_Q(y+\ln
x)\rangle+\langle n_Q(y+\ln (1-x))\rangle-\langle n_G(y)\rangle],\\
\label{eq.14}
\langle n_Q(y)\rangle '&=&\int^1_0
dx\gamma_0^2 K^G_Q(x)[\langle n_G(y+\ln x)\rangle+\langle
n_Q(y+\ln (1-x))\rangle-\langle n_Q(y)\rangle ],
\end{eqnarray}
in which $n_i$ has the meaning of the parton mutiplicity within a jet generated by a parton of type $i=G,Q$.

Using the asymptotic ansatz for the mean multiplicities \cite{basics,Dokshitzer:1993dc}:
\begin{equation}\label{eq.15}
\langle n_G\rangle = e^{\gamma y}, \ \ \ \ \langle n_Q\rangle=\frac{1}{r}e^{\gamma y},
\end{equation} 
which is valid at very high energies, we get from \eqref{eq.12,13} and \eqref{eq.14}
\begin{eqnarray}\label{eq.16}
\gamma&=&\int^1_0dx\gamma_0^2\left[K^G_G(x)[x^{\gamma}+(1-x)^{\gamma}-1]+\frac{n_f}{r}K^Q_G(x)\left[x^{\gamma}+(1-x)^{\gamma}-r\right]\right],\\
\gamma&=&\int^1_0dx \gamma_0^2
K^G_Q(x)\left[rx^{\gamma}+(1-x)^{\gamma}-1\right]\label{eq.16.2}
\end{eqnarray}
respectively. By doing that, we have obtained  two algebraic equations instead the original integro-differential ones. The only variable is the medium characterized by $f_{med}$, and we have a system of two equations with three unknown parameters. Fixing one of them we can solve the system analytically. We study two cases:
\begin{itemize}
\item We fix $\gamma$ as done in \cite{dremin2} and we obtain the dependence of $\alpha_s$ and $r$ with $f_{med}$.
\item We fix $\alpha_s$ and study the variation of $\gamma$ and $r$ with $f_{med}$.
\end{itemize}

\begin{figure}[h]
\begin{center}
\resizebox{0.6\columnwidth}{!}{\includegraphics{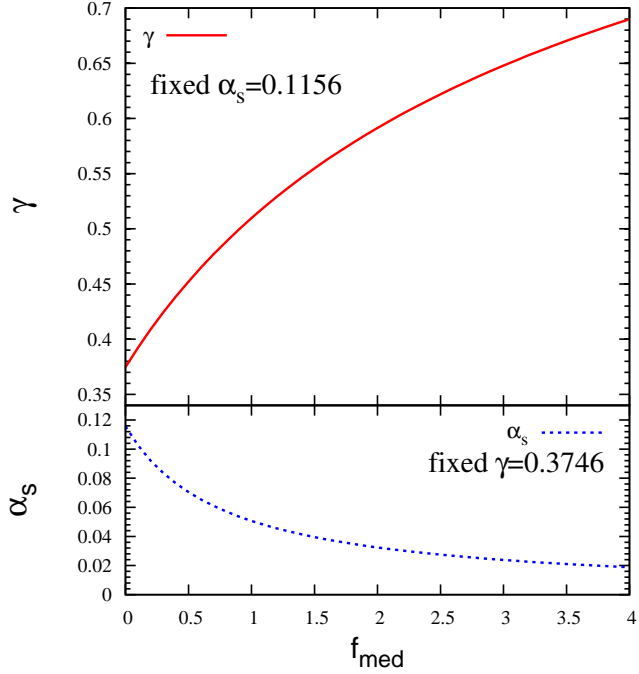}}
\end{center}
\caption{Top: evolution of $\gamma$ with $f_{med}$ for fixed $\alpha_s=0.1156$. Bottom: evolution of  $\alpha_s$ with $f_{med}$ for fixed $\gamma=0.3746$ \cite{dremin2}.}
\label{fig1}
\end{figure}

\begin{figure}[h]
\begin{center}
\resizebox{0.6\hsize}{!}{\includegraphics{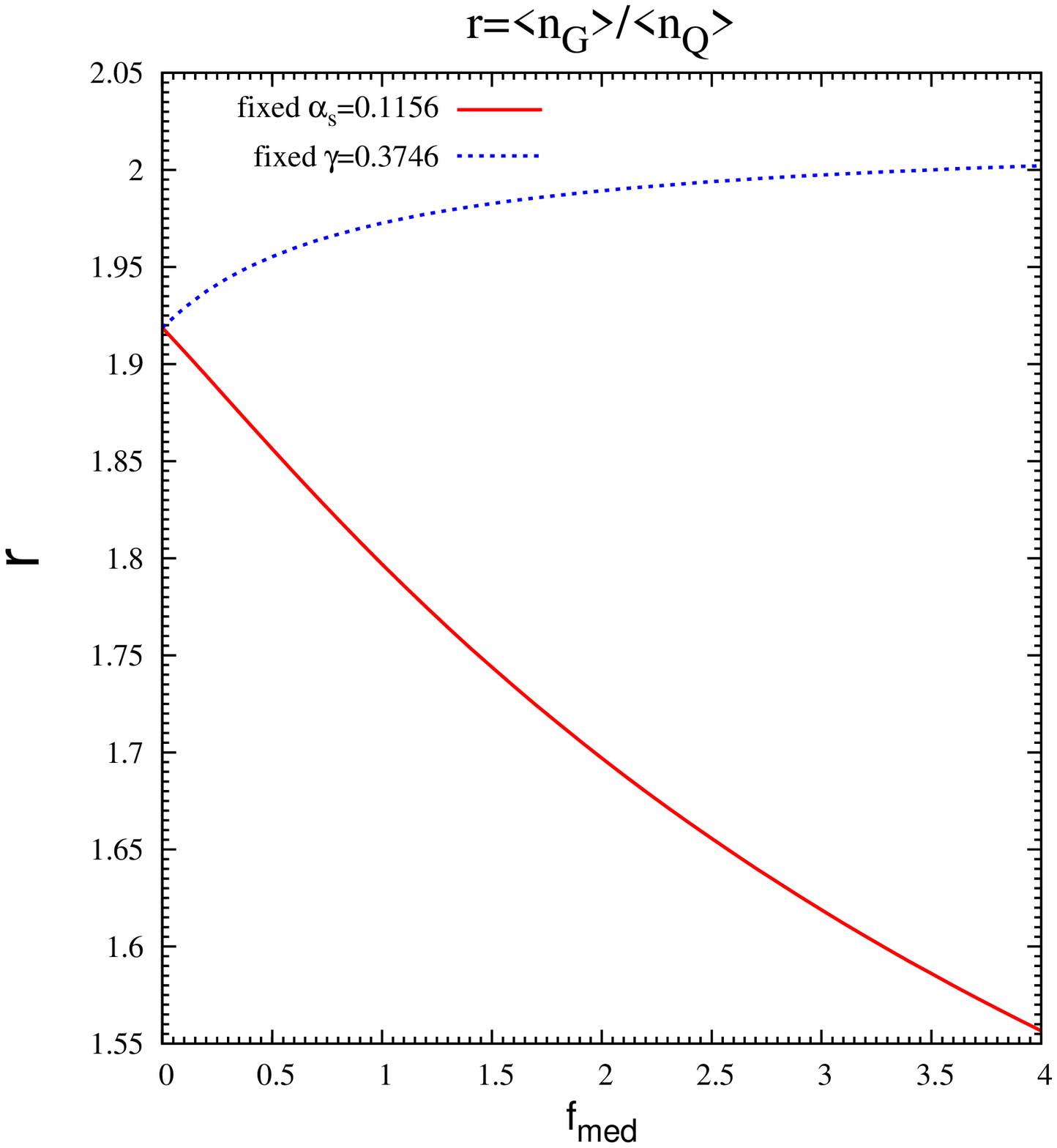}}
\end{center}
\caption{Ratio between gluon and quark mean multiplicities versus $f_{med}$, for both cases of fixed $\alpha_s=0.1156$ (red solid line) and fixed $\gamma=0.3746$ \cite{dremin2} (blue dotted line).}
\label{fig2}
\end{figure}

The results are shown in Figs.~\ref{fig1} and ~\ref{fig2}.  We fix the values of $\gamma$ and $\alpha_s$ in order to agree, for $f_{med}=0$,  with the results in \cite{dremin2}.

When $\gamma$ is set to be medium independent, which with the ansatz \eqref{eq.15} means that the parton mean multiplicity does not change when the jet travels through a medium of increasing density, a decreasing behavior of $\alpha_s$ is found, with $\alpha_s\to 0$ when $f_{med}\to\infty$.
In Fig.~\ref{fig1}bottom we indeed see a very drastic decrease of $\alpha_s$ in a range of $f_{med}$ from $0$ to $4$. As $f_{med}$ is expected to be proportional to $\hat{q}\propto T^3$ \cite{BorghiniWiedemann,Baier:2002tc}, we find a large decrease in the coupling constant for a small increase in temperature.
The ratio of gluon- and quark-jet mean multiplicities, $r$, increases with the medium (Fig. \ref{fig2}), with the large $f_{med}$ limit
\begin{equation}
r\mathop{\to}\limits_{f_{med}\to\infty}\gamma\left[[\gamma_E+\Psi(\gamma)]\left(1-\frac{N_c}{C_F}\right)+\frac{1}{\gamma}\right]\simeq 2,
\end{equation}
meaning that even in the medium gluon jets are still more active than quark jets in producing secondaries.

In the second case we analyze, we fix $\alpha_s$ and calculate the dependence of the anomalous dimension $\gamma$ with the medium. As shown in Fig.~\ref{fig1}top, it increases. The large $f_{med}$ limit is
\begin{equation}
\gamma\mathop{\to}\limits_{f_{med}\to\infty}1.
\end{equation}
Thus, the mean multiplicity of partons increases with the presence of a medium.
When we calculate $r$, we find a decreasing behavior (Fig. \ref{fig2}) with an asymptotic value of 1 when $f_{med}\to\infty$. So the quark and the gluon jets tend to produce secondary partons in the same amount when traversing a medium. This result suggests that for large medium effects, multiplicity is dominated by the gluon piece. It agrees with the behavior observed in \cite{dremin2} of $r$ decreasing with increasing $\gamma$. But
it does not coincide with the values $r=N_c/C_F=9/4$, $\gamma=\sqrt{\gamma_0^2 (1+f_{med})}$ obtained when taking just the infrared-divergent parts of the splitting kernels. It is actually driven by the non-infrared divergent pieces which cancel for $x\to 0$ in \eqref{eq.16} and \eqref{eq.16.2} and, at first glance, it looks dubious. It might be due to the ansatz \eqref{eq.15}. We leave this point for future investigations.

After obtaining the parameters by solving the equations ${\cal O}(z)$, we substitute them in the equations ${\cal O} (z^2)$.

\subsection{Second order equations: ${\cal O}(z^2)$}

Collecting terms with powers $z^2$ in both sides of equations \eqref{eq.9}  and \eqref{eq.9.1} we get:
\begin{eqnarray}\label{eq.18}
\frac{Q_2}{2}(\langle n_Q(y)\rangle^2) '&=&\int^1_0
dx\gamma_0^2K^G_Q(x)\\ \nonumber
&\times&[\frac{1}{2}\langle n_G(y+\ln x)\rangle^2 F_2+\frac{1}{2}\langle
n_Q(y+\ln (1-x))\rangle^2 Q_2\\ \nonumber 
&+&\langle n_G(y+\ln x)\rangle\langle n_Q(y+\ln(1-x))\rangle-\frac{1}{2}\langle n_Q(y)\rangle^2 Q_2],
\end{eqnarray}
\begin{eqnarray}\label{eq.17}
\frac{F_2}{2}(\langle n_G(y)\rangle^2) '&=&\int^1_0
dx\gamma_0^2 K^G_G(x)\frac{1}{2}\\ \nonumber
&\times &[\left(\langle n_G(y+\ln x)\rangle^2 +\langle
n_G(y+\ln (1-x))\rangle^2 \right)F_2 \\ \nonumber
&+&
\langle n_G(y+\ln x)\rangle\langle n_G(y+\ln(1-x))\rangle-\frac{1}{2}\langle n_G(y)\rangle^2 F_2]\\ \nonumber
&+&n_f\int^1_0
dx\gamma_0^2 K^Q_G(x)\frac{1}{2}[\left(\langle n_Q(y+\ln x)\rangle^2+\langle
n_Q(y+\ln (1-x))\rangle^2\right) Q_2\\ \nonumber
&+&\langle n_Q(y+\ln x)\rangle\langle n_Q(y+\ln(1-x))\rangle-\frac{1}{2}\langle n_Q(y)\rangle^2 F_2]
\end{eqnarray}
for the second order normalized factorial moments for quarks ($Q_2$)  and gluons ($F_2$) respectively. Note that in MLLA the second order normalized factorial moments are $y$-independent.

Introducing our ansatz \eqref{eq.15} for the mean multiplicities, these equations read
\begin{eqnarray}
 \label{eq.20}
2Q_2\gamma&=&\int^1_0 dx\gamma_0^2K_Q^G(x)\left( r^2x^{2\gamma}F_2+(1-x)^{2\gamma}Q_2+2rx^{\gamma}(1-x)^{\gamma}-Q_2\right),\\
\label{eq.19}
2\gamma F_2 &=&\int^1_0
dx\gamma_0^2K^G_G(x)\left(x^{2\gamma}F_2+(1-x)^{2\gamma}F_2+x^{\gamma}(1-x)^{\gamma}-F_2\right)\\ \nonumber
&+&\frac{n_f}{r^2}\int^1_0
dx\gamma_0^2K^Q_G\left(x^{2\gamma}Q_2+(1-x)^{2\gamma}Q_2+x^{\gamma}(1-x)^{\gamma}-r^2F_2\right).
\end{eqnarray}

\begin{figure}[h]
\begin{center}
\resizebox{0.6\hsize}{!}{\includegraphics{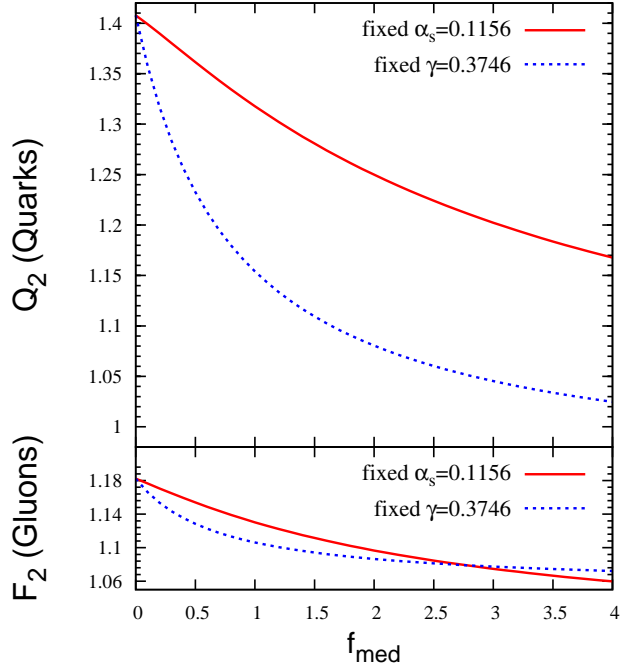}}
\end{center}
\caption{Top: second order normalized factorial moment $Q_2$ for quarks versus $f_{med}$. Bottom:  second order normalized factorial moment $F_2$ for gluons versus $f_{med}$. The line convention is the same as in Fig. \ref{fig2}.}
\label{fig3}
\end{figure}
As we know the analytic expression of $\alpha_s(f_{med})$ in the case of fixed $\gamma$ or $\gamma(f_{med})$ for the case of fixed $\alpha_s$, and $r(f_{med})$, the only unknowns are $Q_2$  and $F_2$.
Solving these system of two algebraic equations we obtain $F_2(f_{med})$ and $Q_2(f_{med})$. We show the results in Fig.~\ref{fig3}, where we can see a decrease in the second order moments for both quark and gluon  jets. The decreasing behavior is found for the two different cases that we analyze.

In order to see what really happens to the multiplicity distribution with the medium, we calculate the dispersion by using the relation given by \eqref{eq.7}:
\begin{eqnarray}\label{eq.17.1}
D^2_Q&=&\langle n_Q\rangle ^2 D^2_{Q,norm}=\frac{1}{ r^2}e^{2\gamma y}(Q_2-1+r e^{-\gamma y}),\\
D^2_G&=&\langle n_G\rangle ^2 D^2_{G,norm} = e^{2\gamma y}(F_2-1+e^{-\gamma y}).
\label{eq.17.2}
\end{eqnarray}
While, as indicated above, in MLLA the second order normalized factorial moments are $y$-independent, this is not the case for the dispersion. On the other hand, the ansatz  \eqref{eq.15} is only valid at large rapidities. Therefore, for the purpose of illustration we choose a large rapidity $y= 10.72$ which roughly corresponds to the total rapidity interval at top RHIC energy.

\begin{figure}[h]
\begin{center}
\resizebox{0.6\hsize}{!}{\includegraphics{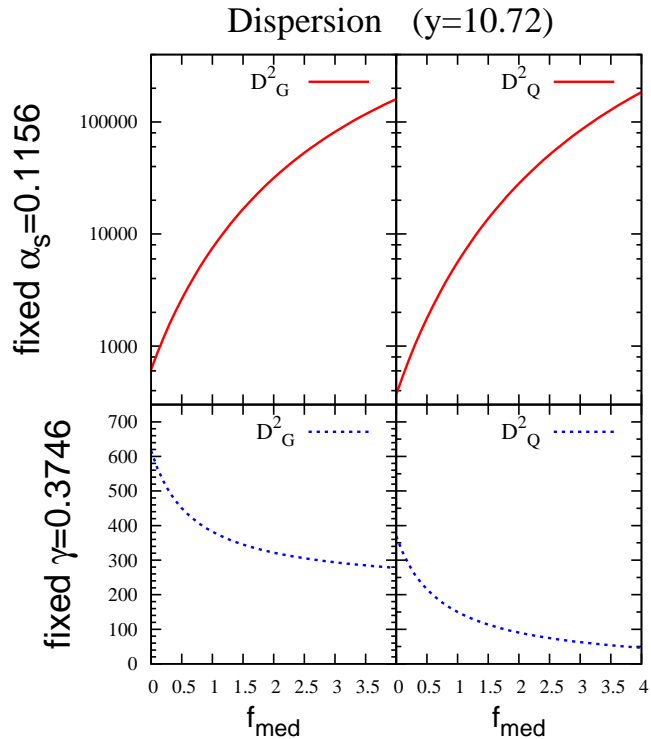}}
\end{center}
\caption{Dispersion for gluons (plots on the left) and quarks (plots on the right) versus $f_{med}$, for both cases of fixed $\alpha_s=0.1156$ (red solid line, upper plots) and fixed $\gamma=0.3746$ \cite{dremin2} (blue dotted line, lower plots), for $y=10.72$.}
\label{fig4}
\end{figure}

In Fig.~\ref{fig4} we show the results obtained for the dispersion of the multiplicity distributions of quark and gluon jets. Now the behaviors in the two cases that we analyze are very different: For fixed $\gamma$ we obtain a decreasing dispersion with increasing $f_{med}$. For fixed $\alpha_s$  an increase in the dispersion of the multiplicity distributions is obtained. The former result is difficult to understand, while the latter makes more sense within the logic that the multiplicity distributions should get wider in medium than in vacuum.

 On the other hand, the normalized dispersions can be directly read from \eqref{eq.17.1} and \eqref{eq.17.2}. For the large rapidity we are examining, the results are similar to those for the second order moments in Fig.~\ref{fig3} shifting the vertical scale down by one unit - the correction due to the last, rapidity-dependent term in the r.h.s. of the equations is smaller than +0.04.

\section{Conclusions}

Analytical estimates of the medium modification of parton branching are most needed, both by their own interest and as a tool for understanding results which come from Monte Carlo simulators \cite{mc}. These medium modifications will hopefully be tested both at RHIC (see the first jet results in \cite{jets}) and at the Large Hadron Collider \cite{lhc} in the near future.

In this work, we have considered the modification of the multiplicity distributions in MLLA due to the presence of a QCD medium. The medium has been introduced though a multiplicative constant
($f_{med}$) in the soft infrared parts of the
kernels of 
QCD evolution equations. Using the asymptotic ansatz \eqref{eq.15} for quark and gluons mean multiplicities, we have studied two cases: fixed $\gamma$ as previously considered in \cite{dremin2}, and 
fixed $\alpha_s$. After analyzing several features of both cases, we have discussed the mutiplicity distributions.
We
find opposite behaviors for the dispersion of the multiplicity 
distributions with increasing $f_{med}$. For fixed $\gamma$ the dispersion decreases, while for fixed $\alpha_s$ it increases.
As future developments, we plan to clarify several points in our analysis, both relaxing the ansatz \eqref{eq.15} and including medium effects in the splitting functions with a more refined treatment than the use of $f_{med}$.

\section*{Acknowledgments}

We thank F. Arleo, S. Jeon, R. P\'erez Ramos, C. A. Salgado and U. A. Wiedemann for useful discussions.
This work has been supported by Ministerio de Educaci\'on
y Ciencia of Spain under project FPA2005-01963, by Xunta de Galicia (Conseller\'{\i}a de Educaci\'on), and by the Spanish Consolider-Ingenio 2010
Programme CPAN (CSD2007-00042). 
NA has been supported by  MEC of Spain under a contract Ram\'on y Cajal, and PQ by MEC of Spain under a grant of the FPU Program; both acknowledge support from Xunta de Galicia through grant PGIDIT07PXIB206126PR.

\end{document}